\documentstyle[12pt]{article}
\newlength{\mytopmargin}
\newlength{\myleftmargin}
\begin{document}

\vspace{4cm}
\noindent
{\large
{\bf NORMALIZATION OF THE WAVEFUNCTION FOR THE \\CALOGERO-SUTHERLAND
MODEL WITH INTERNAL \\ DEGREES OF FREEDOM}}
\vspace{5mm}

\noindent
P.J.Forrester\footnote{email: matpjf@maths.mu.oz.au; supported by the ARC}\\
\noindent
Department of Mathematics, University of Melbourne,Parkville,Victoria
3052,Australia
\vspace{1cm}

\small
\begin{quote}
The exact normalization of a multicomponent generalization of the ground state
wavefunction of the Calogero-Sutherland model is conjectured. This result is
obtained from a conjectured generalization of Selberg's $N$-dimensional
extension of the Euler beta integral, written as a trigonometric integral. A
new proof of the Selberg integral is given, and the method is used to provide
a proof of the multicomponent generalization in a special two-component case.
\end{quote}
\normalsize
\vspace{1.5cm}
\noindent
{\bf 1. INTRODUCTION}
\vspace{5mm}

\noindent
The $1/r^2$ quantum many body system (Calogero-Sutherland model) is the
subject of much present day interest due to its connection with quantum
chaos [1] and fractional statistics [2,3]. The development of these
applications
has been greatly assisted by the recent discovery [4,3] of mathematical methods
which provide the exact evaluation of ground state correlations, both static
and dynamic. These exact calculations rely on $N$-dimensional integration
formulas, which are generalizations of Selberg's [5] $N$-dimensional extension
of the beta integral:
\setcounter{equation}{0}
\renewcommand{\theequation}{1.\arabic{equation}}
\begin{eqnarray}
S(N,\lambda_1,\lambda_2;\lambda) & := &
\left ( \prod_{l=1}^N \int_0^1 dt_l \, t_l^{\lambda_1}(1-t_l)^{\lambda_2}
\right ) \prod_{1 \le j < k \le N} |t_k - t_j|^{2 \lambda} \nonumber \\
& = & \prod_{j=0}^{N-1} {\Gamma (\lambda_1 + 1 + j\lambda)
\Gamma (\lambda_2 + 1 + j\lambda)\Gamma(1+j\lambda) \over
\Gamma (\lambda_1 + \lambda_2 + 2 + (N + j-1)\lambda) \Gamma (1 + \lambda )}
\end{eqnarray}

Generalizations of the Calogero-Sutherland Hamiltonian to include internal
 degrees of freedom of the particles have recently been formulated [6].
These models are of interest in condensed matter physics because of their
relationship with quantum lattice models, notably the $1/r^2$ exchange
$t-J$ Hamiltonian [7]. However, from the viewpoint of exact calculations, the
theoretical development of these models is not as advanced as that of the
original model. In particular, the analogue of the Selberg integral (1.1) for
the exact multicomponent ground state wavefunction [6]
\begin{eqnarray}
\lefteqn{|\psi_0(\{z_j^{(\alpha)}\}_{\alpha = 1,\dots,p \atop
j=1,\dots,N_\alpha},
\{w_j\}_{j=1,\dots,N_0})|^2 }\nonumber \\
& = & \prod_{\alpha=1}^p \prod_{1 \le j < k \le N_{\alpha}}
|z_j^{(\alpha)} - z_k^{(\alpha)}|^{2 (\lambda + 1)}
\prod_{1 \le j' < k' \le N_0}|w_{k'}-w_{j'}|^{2\lambda}\nonumber \\
& &\times \prod_{\alpha=1}^p\prod_{\beta=1 \atop \beta \not= \alpha}^p
\prod_{j=1}^{N_\alpha}\prod_{k=1}^{N_\beta}
|z_j^{(\alpha)} - z_k^{(\beta)}|^{2 \lambda }\:
\prod_{\alpha=1}^p\prod_{j=1}^{N_\alpha}\prod_{j'=1}^{N_0}
|z_j^{(\alpha)} - w_{j'}|^{2 \lambda }
\end{eqnarray}
where
$$
z_j^{(\alpha)} := e^{2 \pi i x_j^{(\alpha)}/L} \qquad \mbox{and} \qquad
w_j : =  e^{2 \pi i y_j/L}
$$
is not known in the existing literature.

This deficiency has motivated us to pursue the task of formulating the
appropriate analogue of the Selberg integral.
We first transform the Selberg integral into an equivalent
form due to Morris [8]:
\begin{eqnarray}
\lefteqn{D(N;a,b,c)}\nonumber \\
&:= & {\rm CT} \prod_{l=1}^N(1-t_l)^a(1-{1 \over t_l})^b
\prod_{1 \le j < k \le N}(1-{t_k\over t_j})^c(1 - {t_j \over t_k})^c \nonumber
\\
& = & \left ( \prod_{l=1}^N \int_{-1/2}^{1/2} d\theta_l \,
e^{\pi i \theta_l (a-b)}|1 + e^{2 \pi i \theta_l}|^{a+b} \right )
 \prod_{1 \le j < k \le N}|e^{2 \pi i \theta_k}-e^{2 \pi i \theta_j}|^{2c}
\nonumber \\
& = & \prod_{l=0}^{N-1} {\Gamma (a+b+1+lc) \Gamma (1+(l+1)c) \over
\Gamma (a+1+lc)\Gamma (b+1+lc) \Gamma (1+c)}
\end{eqnarray}
where CT denotes the constant term in the Laurent expansion. The constant
term is well defined when the function is a Laurent polynomial: this requires
$a+b,c \in Z_{\ge 0}, a,b \in Z$. However the Fourier integral in (2.3)
defines an analytic function for ${\rm Re}(a+b)>1, {\rm Re}(c) \ge 0$
(at least) and is evaluated by the product of gamma functions whenever
it is defined.
 By using a combination of
analytical and numerical methods we are able to conjecture the
value of (1.3) when the product of differences is replaced
by (1.2), and the one-body factors extended over all particle
coordinates. We are able to provide analytic checks on the validity of
our conjecture. In fact, in the case $p=1, \lambda = 1$ these checks suffice to
prove
the conjecture. Our results include the conjectured exact normalization
of the wavefunction (1.2). They can also be cast in a form which directly
generalizes the Selberg integral (1.1). Furthermore, they also include the
conjectured exact
normalization of the ground state wavefunction
\begin{eqnarray}
\lefteqn{|\psi_0^{(h)}(\{x_j^{(\alpha)}\}_{\alpha = 1,\dots,p \atop
j=1,\dots,N_\alpha},
\{y_j\}_{j=1,\dots,N_0})|^2 }\nonumber \\
& = &\prod_{\alpha=1}^p\prod_{j=1}^{N_\alpha}e^{-x_j^{(\alpha)2}/2}
 \prod_{\alpha=1}^p \prod_{1 \le j < k \le N_{\alpha}}
|x_j^{(\alpha)} - x_k^{(\alpha)}|^{2 (\lambda + 1)} \nonumber \\
& & \times \prod_{j'=1}^{N_0}e^{-y_j^2/2}
\prod_{1 \le j' < k' \le N_0}|y_{k'}-y_{j'}|^{2\lambda}\nonumber \\
& & \times \prod_{\alpha=1}^p\prod_{\beta=1 \atop \beta \not= \alpha}^p
\prod_{j=1}^{N_\alpha}\prod_{k=1}^{N_\beta}
|x_j^{(\alpha)} - x_k^{(\beta)}|^{2 \lambda }\:
\prod_{\alpha=1}^p\prod_{j=1}^{N_\alpha}\prod_{j'=1}^{N_0}
|x_j^{(\alpha)} - y_{j'}|^{2 \lambda }
\end{eqnarray}
for the system in a harmonic
well.
\vspace{1cm}

\noindent
{\bf 2. EQUIVALENCE BETWEEN SELBERG-TYPE INTEGRALS AND\\ SOME FOURIER
INTEGRALS}

\vspace{.5cm}
\noindent
{\bf 2.1 The inter-relationship}

\noindent
To perform a numerical investigation of the value of the Selberg integral (1.1)
or any generalizations, it is convenient to first transform the Selberg-type
integral
into an equivalent Fourier integral. For this purpose we use the following
lemma.
\vspace{.2cm}

\noindent
{\bf Lemma 1}

\noindent
Let $f(t_1,\dots,t_N;\{ p \})$ be a Laurent polynomial in $t_1, \dots,t_N$,
with
$\{ p \}$ as parameters. For ${\rm Re} (\epsilon)$ large enough so that the
r.h.s. exists
$$
\left ({\pi \over \sin \pi \epsilon} \right )^N
\left ( \prod_{l=1}^N \int_{-1/2}^{1/2} d\theta_l \, e^{2 \pi i \theta_l
\epsilon}
\right )
f(-e^{2 \pi i \theta_1},\dots,-e^{2 \pi i \theta_N};\{ p \}) \hspace{1cm}
$$
$$
= \left ( \prod_{l=1}^N \int_0^1 dt_l \, t_l^{-1+\epsilon}
\right )f(t_1,\dots,t_N;\{ p \})
\eqno (2.1)
$$
\vspace {.2cm}

This result follows immediately from term-by-term integration of the Laurent
polynomial.
Note that for $\epsilon$ an integer the Fourier integral is equal to
$$
{\rm CT}_{\{t_1,\dots,t_N\}} \prod_{l=1}^N t_l^\epsilon \:
f(-t_1,\dots,-t_N;\{ p \})
\eqno (2.2)
$$

{}From this lemma we see that the Selberg integral with $\lambda_1$ arbitrary
and $\lambda_2, \lambda$ non-negative integers is equivalent to the Fourier
integral in Morris's integral (1.3) with $a-b$ arbitrary and $a+b, c$
non-negative
integers.

\vspace{.2cm}
\noindent
{\bf 2.2 Numerical evaluation}

\noindent
For $\epsilon$ an integer the Fourier integral in (2.1) can be
 computed by exact numerical integration.
Thus, whenever
$$
g(x) = \sum_{n_1 = -p_1}^{p_1}\dots \sum_{n_N = -p_N}^{p_N}
a_{n_1,\dots,n_N} e^{2 \pi i (x_1n_1 + \dots x_Nn_N)}
$$
we have
$$
\int_0^1 dx_1 \dots \int_0^1 dx_N \, g(x_1,\dots,x_N) =
{1 \over M_1}\sum_{n_1 = 0}^{M_1} \dots {1 \over M_N}\sum_{n_N = 0}^{M_N}
g(n_1/M_1, \dots, n_N/M_N)
\eqno (2.3)
$$
provided $M_l > p_l$ $(l=1,\dots,N)$. This result follows by term-by-term
integration and summation of the Fourier series for $g(x)$. In both cases only
the $n_j = 0$ term $(j=1,\dots,N)$ of $g(x)$ remains.

We will use the formula (2.3) below to provide numerical data on the evaluation
of some generalizations of (1.3).

\vspace{.2cm}
\noindent
{\bf 2.3 Analytic properties of the Fourier integrals}

\noindent
Denote the Fourier integral on the l.h.s. of (2.1) by $I(\epsilon;\{p\})$.
Suppose furthermore that $f$ is real when each $\theta_j$ is real so that
$$
f(-e^{2 \pi i \theta_1},\dots,-e^{2 \pi i \theta_N};\{ p \})=
f(-e^{-2 \pi i \theta_1},\dots,-e^{-2 \pi i \theta_N};\{ p \})
\eqno (2.4)
$$
Some immediate properties of $I(\epsilon;\{p\})$ are \\
(i) $I(\epsilon;\{p\})$ is an entire function of $\epsilon$. \\
(ii) $I(\epsilon;\{p\}) =I(-\epsilon;\{p\})$ (this property requires (2.4)). \\
(iii) $I(\epsilon;\{p\}) =\Big ( (\sin \pi \epsilon)/\pi
 \Big )^N J(\epsilon;\{p\})$, where $J(\epsilon;\{p\})$, which is given by the
r.h.s. of (2.1) for ${\rm Re} (\epsilon)$ large enough, is a rational function
of $\epsilon$.

Consider further property (iii). Suppose in fact that $J(\epsilon;\{p\})$
is the reciprocal of a polynomial:
$$
J(\epsilon;\{p\})= {k(\{p\}) \over
\prod_{j=1}^{M(\{p\})}(\epsilon + n_j(\{p\}))^{q_j(\{p\})} }
\eqno (2.5)
$$
Then $q_j(\{p\})$ is the  maximum number of times the integer
$n_j(\{p\})$  occurs as a power in a single term of the Laurent
expansion of $f$. We will see below that the Selberg-type integrals
related to the wavefunction (1.2) have the property (2.5).

In the cases that (2.5) holds, a conjectured evaluation of (2.1) of the form
$$
\Big ( (\sin \pi \epsilon)/\pi
 \Big )^N  \alpha(\{p\}) A_N(\epsilon;\{p\})
\eqno (2.6)
$$
can be proved to be correct up to the multiplicative function of the parameters
$\alpha(\{p\})$ by an application of Liouville's theorem. Thus we need to
show that $1/ A_N(\epsilon;\{p\})$ is a polynomial in $\epsilon$ and to specify
the positions and orders of its zeros. Then we need to show that the positions
 of the poles of $ J_N(\epsilon;\{p\})$ coincide with these zeros, and their
order
is no greater than that of the corresponding zero.
Finally, we need to check that $ J_N(\epsilon;\{p\})$ is the reciprocal of
a polynomial by calculating its $|\epsilon|\rightarrow \infty$ behaviour.

Let us illustrate this method to prove that for $a+b := 2r, c \in Z_{\ge 0}$,
the Fourier integral in (1.3) as a function of $a-b := 2\epsilon$ is equal
to the product of gamma functions in (1.3) up to a multiplicative function
of $N, c $ and $r$. For this problem we have
$$
f(t_1,\dots,t_N;\{ p \}) = \prod_{l=1}^N(1 - t_l)^r(1 - 1/t_l)^r
\prod_{1 \le j < k \le N} (1 - t_k/t_j)^c(1 - t_j /t_k)^c
\eqno (2.7)
$$
and using the functional equation for the gamma function
$$
\Gamma (z) \Gamma (1 -z) = {\pi \over \sin \pi z}
$$
we see that the product of gamma functions in (1.3) can be written in the form
of (2.6) with
$$
A_N(\epsilon;\{p\})=\prod_{l=0}^{N-1}{\Gamma (\epsilon - r -lc) \over
\Gamma (\epsilon + r + 1 + lc) }
\eqno (2.8)
$$

Consider (2.8). Since $r$ and $c$ are integers we see from the recurrence
of the gamma function
$$
\Gamma (z+1) = \Gamma (z)
$$
that $A_N(\epsilon;\{p\})$ is the reciprocal of a polynomial, with poles of
order $j$ $(j=1,\dots,N-1)$ at
$$
|\epsilon| = r+1+(N-j-1)c, r+2+(N-j-1)c, \dots, r+c+(N-j-1)c
\eqno (2.9)
$$
and poles of order $N$ at
$$
|\epsilon|=0,\dots,r
\eqno (2.10)
$$
The order of the polynomial is equal to
$$
\sum_{n=1}^N l_n n
$$
where $l_n$ is the number of poles of order $n$. Since $l_n = 2c$ $(n=1,\dots,
N-1)$ and $l_N = 2r+1$ the order is thus
$$
cN(N-1) + N(2r+1)
\eqno (2.11)
$$

Now consider the r.h.s. of (2.1) with $f$ given by (2.7). Then (2.1) can be
written in the form (2.5), where in general $k(\{p\})$ may be a polynomial
in $\epsilon$. To prove that $A_N(\epsilon;\{p\})$
is given by (2.8) we need to show three features: \\
(a) $|n_j(\{p\})|$ in (2.5) is no greater than $r+c(N-1)$. \\
(b) For $|n_j(\{p\})|$ given by the r.h.s. of (2.9) and (2.10),
 $q_j(\{p\})$ is less than or equal to the order of
the corresponding pole in (2.9) and (2.10). \\
(c) For large-$|\epsilon|$, $J_N(\epsilon;\{p\})$ has an inverse power law
decay with exponent (2.11).

To check (a), we see from (2.7) that the largest power of say $t_1$
is $r$ in the
product over $l$ and $c(N-1)$ in the product over $j$ and $k$. It is thus
$r+c(N-1)$ in total, which proves (a) for $n_j(\{p\})$ positive. Since (2.7) is
also
unchanged by the replacement $t_j \mapsto 1/t_j$ $(j=1,\dots,N)$, the same is
true for $n_j(\{p\})$ negative.

Consider statement (b). With $q^{(N)}(n)$ denoting the maximum number of times
the exponent $n$ occurs in any term of the Laurent expansion of (2.7),
statement (b) says
$$
q^{(N)}(n) \le N \qquad \mbox{for} \qquad |n| \le r
\eqno (2.12a)
$$
$$
q^{(N)}(n) \le N-j \qquad \mbox{for} \qquad |n|=r+(j-1)c+\nu
\eqno (2.12b)
$$
where $j=1,\dots,N-1$ and $\nu = 1,\dots,c$. Since there are only $N$ variables
$q^{(N)}(n) \le N$ for any $n$ so (2.12a) is true. To prove (2.12b) consider
the explicit formula (2.7) for $f$. Since the first two products can create all
powers $|n| \le r$ in each variable independently, we see (2.12b) is
equivalent to saying
$$
\hat{q}^{(N)}(n) \le N-j \qquad \mbox{for} \qquad |n|=(j-1)c+\nu
\eqno (2.13)
$$
where $\hat{q}^{(N)}(n)$ denotes the maximum number of times the exponent
$n$ occurs in the Laurent expansion of
$$
\prod_{1 \le j < k \le N} (1 - t_k/t_j)^c(1 - t_j /t_k)^c \hspace{4cm}
$$
$$
=(-1)^{cN(N-1)/2} \prod_{j=1}^N t_j^{-c(N+1)+2cj}
\prod_{1 \le j < k \le N} (1 - t_k/t_j)^{2c}
\eqno (2.14)
$$
Writing
$$
{t_j \over t_k} = {t_j \over t_{j+1}} {t_{j+1} \over t_{j+2}} \dots
{t_{k-1} \over t_k}, \qquad j < k
\eqno (2.15)
$$
we see by expanding the last product in (2.14) that the terms in the Laurent
expansion of (2.14) are of the form [9]
$$
\prod_{j=1}^N t_j^{-c(N+1) + 2cj + n_{j+1} - n_j},
\eqno (2.16)
$$
where $n_1 = n_{N+1} = 0$ and $n_j \ge 0$ for each $j=2,\dots,N$. Equivalently,
setting
$$
n_{p+1} = c(N-p)p + m_{p+1}, \qquad m_{p+1} \ge -c(N-p)p
\eqno (2.17)
$$
for each $p=1,\dots,N-1$ we have that all terms in the Laurent expansion of
(2.16) are of the form
$$
t_N^{-m_N} t_{N-1}^{m_N - m_{N-1}} \dots t_p^{m_{p+1} - m_p} \dots t_1^{m_2}
\eqno (2.18)
$$

We want to determine the maximum number of exponents in (2.18) which can take
the
value $(j-1)c + \nu$. Since (2.14) is a symmetrical function of all the
variables
we can suppose that the $k$ variables $t_N,t_{N-1},\dots,t_{N+1-k}$ have
exponent $(j-1)c + \nu$. Then from (2.18) we require
$$
m_{p+1} = -((j-1)c + \nu)(N-p), \qquad p=N-1, \dots, N-k
\eqno (2.19)
$$
Combining (2.19) with the inequality in (2.16) gives
$$
(j-1)c + \nu \le cp \le c(N-k)
\eqno (2.20)
$$
which, since $1 \le \nu \le c$ implies
$$
k \le N-j
\eqno (2.21)
$$
The inequality (2.21) is precisely the statement (2.13) with $n=(j-1)c + \nu$.
Also, since (2.14) is unchanged by replacing all variables by their
reciprocals,
(2.21) also establishes (2.13) for the remaining case $n=-((j-1)c + \nu)$.

To check (c) we note from (2.7) and (1.3) that
$$
J_N(\epsilon;\{p\})=(-1)^{rN + cN(N-1)/2}\left (\prod_{l=1}^N
\int_0^1dt_l \,t_l^{-1+\epsilon-r-(N-1)c}(1-t_l)^{2r} \right )
$$
$$
\times \prod_{1 \le j < k \le N} (t_k - t_j )^{2c}
\eqno (2.22)
$$
The change of variables
$$
t_l = e^{-s_l} \qquad \mbox{then} \qquad s_l \mapsto s_l/\epsilon
$$
gives the large-$|\epsilon|$ asymptotic behaviour
$$
J_N(\epsilon;\{p\}) \: \sim \: {h_N(\{ p \}) \over \epsilon^{cN(N-1) +
N(2r+1)}}
\eqno (2.23)
$$
with
$$
h_N(\{ p \}):=(-1)^{rN + cN(N-1)/2}\left (\prod_{l=1}^N
\int_0^\infty ds_l \,s_l^{2r}e^{-s_l} \right )
\prod_{1 \le j < k \le N} (s_k - s_j )^{2c}
$$
which is precisely the inverse power law decay with exponent (2.11) required by
(c).

By checking (a)-(c) we have shown, by Liouville's theorem that the Fourier
integral in (1.3) as a function of
$a-b:= 2\epsilon$ is evaluated by the product of gamma functions in (1.3), up
to multiplicative terms independent of $\epsilon$. \\
Remark: The multiplicative function of $N,c$ and $r=(a+b)/2$ in (1.3)
undetermined
by the above can readily be calculated (see the final paragraphs of section
3.1 for the method). We have thus provided a new proof of Morris's integral
and consequently of the Selberg integral.

\vspace{.2cm}

\noindent
{\bf 2.4 Notation}

\noindent
In the remainder of this paper we will consider the generalization of (1.3)
\renewcommand{\theequation}{2.24}
\begin{eqnarray}
\lefteqn{D_p(N_1,\dots,N_p;N_0;a,b,\lambda)} \nonumber \\
& := & \left ( \prod_{\alpha = 1}^p \prod_{j=1}^{N_\alpha} \int_{-1/2}^{1/2}
 d x_j^{(\alpha)} \,
e^{\pi i x_j^{(\alpha)} (a-b)}|1 + e^{2 \pi i x_j^{(\alpha)}}|^{a+b} \right )
\nonumber \\
& & \times \left ( \prod_{l=1}^{N_0} \int_{-1/2}^{1/2} dy_l \, e^{\pi i (a-b)
y_l}
|1 + e^{2 \pi i y_l}|^{a+b} \right )
\psi_0(\{e^{2 \pi ix_j^{(\alpha)}}\}_{\alpha = 1,\dots,p \atop
j=1,\dots,N_\alpha},
\{e^{2 \pi i y_j}\}_{j=1,\dots,N_0}) \nonumber \\
\end{eqnarray}
where $\psi_0$ is given by (1.2). We notice that
$$
D_1(N;0;a,b,c-1) = D_0 (N; a,b,c) = D(N; a,b,c)
\eqno (2.25)
$$
where $ D(N; a,b,c)$ is the integral in (1.3).

\vspace{.5cm}
\noindent
{\bf 3. THE CASE \mbox{\boldmath$p=1$}}

\vspace{.2cm}
\noindent
In the case $p=1, a=b=0$, the exact evaluation of (2.24) is known analytically
 [10]:
\renewcommand{\theequation}{3.1}
\begin{eqnarray}
\lefteqn{D_1(N_1;N_0;0,0,\lambda)} \nonumber \\
& = & {\Gamma( \lambda N_0 + (\lambda + 1)N_1 + 1) \Gamma (N_1 + 1)
\over \Gamma (1 + \lambda)^{N_0} \Gamma (2 + \lambda)^{N_1} \left (
1 + {\scriptstyle \lambda N_0 \over \lambda + 1} \right )_{N_1}}
\end{eqnarray}
where $(a)_n := a(a+1) \dots (a+n-1)$.

For general $a$ and $b$ the method used in [10] to prove (3.1) does not appear
to
be applicable. We thus resorted to the numerical approach outlined in Section
2.2.

\vspace{.2cm}
\noindent
{\bf 3.1 The case  \mbox{\boldmath$\lambda = 1$}}

\noindent
With $\lambda = 1, N_0 =1$ and 2, and various values of $a$ and $b$, we found
by sequentially increasing $N_1$ that our data fitted the  following forms
\begin{eqnarray*}
D_1(N_1;1;1,1,1)& = &2 \prod_{j=0}^{N_1-1} (j+1)(2j+3) \\
D_1(N_1;1;2,1,1)& =  &3 \prod_{j=0}^{N_1-1} (j+1)(2j+4) \\
D_1(N_1;1;2,2,1)& = &6 \prod_{j=0}^{N_1-1} {(j+1)(2j+4)(2j+5) \over 2j+3}\\
D_1(N_1;2;1,1,1) &= &6 \prod_{j=0}^{N_1-1} (j+1)(2j+4)\\
D_1(N_1;2;2,1,1) &= &12 \prod_{j=0}^{N_1-1} (j+1)(2j+5)\\
D_1(N_1;2;2,2,1) &= &40 \prod_{j=0}^{N_1-1} {(j+1)(2j+5)(2j+6) \over 2j+4}
\end{eqnarray*}
valid for $N_1 \ge 0$ (when $N_1 = 0$ the products are taken as unity).

To help fit this data into an analytic form we note from Morris's integral
(1.3)
in the case $c=2$ and $a \in Z_{\ge 0}$ and $b \in Z^+$  that
$$
D_1(N_1;0;a,b,1) = \prod_{j=0}^{N_1 - 1}
{(j+1)(2j+a+1)(2j+a+2) \dots (2j+a+b) \over (2j+2)(2j+3) \dots (2j+b)}
\eqno (3.2)
$$
We see that the above data fits a similar form:
\renewcommand{\theequation}{3.3}
\begin{eqnarray}
\lefteqn{D_1(N_1;N_0;a,b,1)} \nonumber \\ & = & f(N_0,a,b)
\prod_{j=0}^{N_1 - 1}
{(j+1)(2j+a+N_0+1)(2j+a+N_0+2) \dots (2j+a+b+N_0) \over
(2j+N_0+2)(2j+N_0+3) \dots (2j+N_0+b)} \nonumber \\
& = &f(N_0,a,b)
\prod_{j=0}^{N_1 - 1} {(j+1) \Gamma (2j+a+b+N_0+1)\Gamma(2(j+1)+N_0) \over
  \Gamma (2j+a+N_0+1)\Gamma (2j+b+N_0+1)}
\end{eqnarray}
To evaluate $ f(N_0,a,b)$ we set $N_1=0$ in (3.3) (the product over $j$ is then
taken as unity) to obtain
$$
f(N_0,a,b) = D_1(0;N_0;a,b,2)=D(N_0;a,b,1)
\eqno (3.4)
$$
where $D(N_0;a,b,1)$ is given by (1.3).

 The resulting conjecture for $D_1$ can
be proved using the method of Section 2.3.
Thus we consider $D_1(N_1;N_0;a,b,1)$ as a function of $a-b :=2 \epsilon$.
We have the following result.

\vspace{.2cm}
\noindent
{\bf Theorem 1}

\noindent
Suppose $r \in Z_{\ge 0}$ and let
\renewcommand{\theequation}{3.5}
\begin{eqnarray}
\lefteqn{f(t_1,\dots,t_{N_0};s_1,\dots,s_{N_1};r)} \nonumber \\
& := & \prod_{l=1}^{N_0} (1 - t_l)^r(1 - 1/t_l)^r
\prod_{l=1}^{N_1}(1 - s_l)^r(1-1/s_l)^r \nonumber \\
& &\times  \prod_{1 \le j < k \le N_0} (1-t_k / t_j)(1-t_j/t_k)
\prod_{1 \le j < k \le N_1} (1-s_k/s_j)^2(1-s_j/s_k)^2 \nonumber \\
& &\times  \prod_{j=1}^{N_0} \prod_{k=1}^{N_1} (1-s_k/t_j)(1-t_j/s_k)
\end{eqnarray}
Then for ${\rm Re}(\epsilon)$ large enough so that the l.h.s. converges,
$$
\left ( \prod_{l=1}^{N_0} \int_0^1 dt_l \, t_l^{-1+\epsilon} \right )
\left ( \prod_{l=1}^{N_1} \int_0^1 ds_l \, s_l^{-1+\epsilon} \right )
f(t_1,\dots,t_{N_0};s_1,\dots,s_{N_1};r)
$$
$$
= A(N_0,N_1,r) \prod_{l=0}^{N_0-1}{\Gamma (\epsilon - r - l) \over
\Gamma ( \epsilon + r + 1 + l) }
\prod_{l=0}^{N_1-1} {\Gamma(\epsilon - r - 2l -  N_0)
\over \Gamma (\epsilon + r + 1 + 2l +  N_0 ) }
\eqno (3.6)
$$
where $A(N_0,N_1,r)$ is independent of $\epsilon$.

\vspace{.2cm}
\noindent
{\bf Proof}

\noindent
Since $r \in Z_{\ge 0}$ we see that the r.h.s. of (3.6) is the reciprocal of a
polynomial in $\epsilon$. Furthermore this polynomial is naturally factored as
two polynomials, one for each of the products. The first reciprocal polynomial
factor is precisely (2.8) with $c=1$ and $N=N_0$. This factor therefore
has poles at (2.9) and (2.10) with the order given therein. The second factor
has poles of order $j \: (j=1,\dots,N_1-1)$ at integers $\epsilon$ satisfying
$$
r+N_0 + 2(N_1 - j) \le |\epsilon| \le r+N_0+2(N_1-j)+1
\eqno (3.7)
$$
and poles of order $N_1$ at
$$
|\epsilon| = 0,1, \dots,r+N_0
\eqno (3.8)
$$

The poles at (3.7) and (3.8) occur at the same points as the poles of the first
factor, plus some additional points. Thus if we let $Q^{(N_0,N_1)}(n)$ denote
the order of the pole of (3.5) at $n,\: (n \in Z)$, we see that
$$
Q^{(N_0,N_1)}(n) = N_0 + N_1, \qquad |n| \le r
\eqno (3.9a)
$$
$$
Q^{(N_0,N_1)}(n)=N_1 + N_0 -j_1, \qquad |n|=r+j_1
\eqno (3.9b)
$$
where $j_1=1,\dots,N_0$,
$$
Q^{(N_0,N_1)}(n)=N_1-j_2, \qquad |n|=r+N_0+2(j_2-1) + \nu_2
\eqno (3.9c)
$$
where $j_2=1, \dots, N_1-1, \: \nu_2=1,2,$ and
$$
Q^{(N_0,N_1)}(n)=0, \qquad \mbox{otherwise}.
\eqno (3.9d)
$$
{}From (3.9) we see that the reciprocal of the r.h.s. of (3.6) is a polynomial
of order
$$
(2r+1)(N_0+N_1) + 2 \sum_{j=0}^{N_0-1} (j+N_1) + 4 \sum_{j=1}^{N_1-1}j
\hspace{4cm}
$$
$$
=(2r+1)(N_0+N_1) + 2N_0N_1 + (N_0-1)N_0 + 2(N_1-1)N_1
\eqno (3.10)
$$

Let $q^{(N_0,N_1)}(n)$ denote the maximum number of times the exponent $n$ can
occur in a term of the Laurent expansion of (3.5). In accordance with the
method
of Section 2.3 we want to show that
$$
q^{(N_0,N_1)}(n) \le Q^{(N_0,N_1)}(n)
\eqno (3.11)
$$

First, for $|n| \le r$ there is nothing to prove as (3.11) reads
$$
q^{(N_0,N_1)}(n) \le N_0 + N_1
$$
which is true by definition. Next consider the statement (3.11) for $|n|$ and
$Q^{(N_0,N_1)}(n)$ given by (3.9b). Since the products over $l$ in (3.5) give
all exponents $n$ of $t_1,\dots,t_{N_0},s_1,\dots,s_{N_1}$ with $|n| \le r$
in each variable independently, and (3.5) is unchanged by replacing each
variable
by its reciprocal, we see that in these cases (3.11) is equivalent to
proving
$$
\tilde{q}^{(N_0,N_1)}(n) \le N_1 + N_0 -j_1, \qquad n=j_1
\eqno (3.12a)
$$
and
$$
\tilde{q}^{(N_0,N_1)}(n)  \le N_1 - j_2, \qquad n= N_0 + 2(j_2 - 1) + \nu_2
\eqno (3.12b)
$$
where $\tilde{q}^{(N_0,N_1)}(n) $ denotes the maximum number of times the
exponent $n$ can occur in the Laurent expansion of
$$
\prod_{1 \le j < k \le N_0} (1-t_k / t_j)(1-t_j/t_k)
\prod_{1 \le j < k \le N_1} (1-s_k/s_j)^2(1-s_j/s_k)^2
$$
$$
\times
\prod_{j=1}^{N_0} \prod_{k=1}^{N_1} (1-s_k/t_j)(1-t_j/s_k)
\eqno (3.13)
$$

Using a confluent form of of the Vandermonde determinant expansion, we have
previously shown [11] that (3.13) is equal to (up to an unimportant $\pm$ sign)
$$
\sum_{P \atop P(2l) > P(2l-1)} \epsilon (P) \sum_{Q=1}^{N!} \epsilon (Q)
\prod_{j=1}^{N_1} s_j^{P(2j)+P(2j-1)-2N_0-N_1-1} (P(2j) - P(2j -1))
$$
$$
\times
\prod_{k=1}^{N_0} t_k^{P(2N_1 + k) + Q(k) - N_0 - N_1 -1}
\eqno (3.14)
$$
where $P$ is a permutation of $\{ 1,2, \dots, 2N_1 + N_0\}$ with parity
$\epsilon (P)$ and $Q$ is a permuation of $\{ 1, \dots, N_0 \}$ with parity
$\epsilon (Q)$.

Let $j_1, k \in \{ 1, \dots, N_0 \}$ and $j \in \{1, \dots, N_1 \}$. We see
from (3.14) that an exponent of $j_1$ in the variable $t_k$ requires
$$
P(2N_1 + k) = j_1 + N_0 + N_1 + 1 - Q(k)
\eqno (3.15a)
$$
and thus
$$
P(2N_1 + k) \in \{ j_1 + N_1 + 1, j_1 + N_1 +2, \dots, {\rm min}(2N_1+N_0,
j_1 + N_1 +N_0) \}
\eqno (3.15b)
$$
For an exponent of $j_1$ in the variable $s_j$ we see from (3.14) that we
require
$$
P(2j) + P(2j-1) = 2N_1 + N_0 + 1 + j_1, \qquad P(2j) > P(2j-1).
\eqno (3.16)
$$

Consider the case $j_1 \le N_1$. Then
$$
{\rm min}(2N_1 + N_0, j_1 + N_1 + N_0) = j_1 + N_1 + N_0
$$
and so the maximum number of solutions of (3.15a) is $N_0$. When (3.15a) has
this maximum number of solutions, (3.16) has solutions for
$$
(P(2j),P(2j-1)) = (j_1+1+N_0+N_1,N_1),\,(j_1+2+N_0+N_1,N_1-1),
\dots,(2N_1+N_0,j_1+1),
$$
thus giving a maximum of $N_1-j_1$ exponents $j_1$ to the variable $t_j$, and
thus a total of
$N_0+N_1 - j_1$ exponents $j_1$ in (3.14). Since decreasing the number of
solutions of (3.15a) by say $a_1$ can give no more than $a_1$ new solutions to
(3.16), we thus have shown that for $j_1 \le N_1$
$$
\tilde{q}^{(N_0,N_1)}(j_1) \le N_1 + N_0 -j_1
\eqno (3.17)
$$

In the cases that $j_1 > N_1$ (which requires $N_0 > N_1$)
$$
{\rm min}(2N_1 + N_0, j_1 + N_1 + N_0) = 2N_1 + N_0
$$
and so the maximum number of solutions of (3.15a) is $N_1 + N_0 - j_1$.
When (3.15a) has this maximum number of solutions, there are no solutions
to (3.16). Arguing as in the sentence including (3.17), we conclude that
(3.17) also holds for $j_1 > N_1$ and thus (3.12a) is true.

For an exponent of $n=N_0 + 2(j_2 - 1) + \nu_2$ in the variable $s_j$, (3.14)
gives that we require (3.16) with $j_1$ replaced by $n$. The maximum number of
solutions occurs with
$$
(P(2j),P(2j-1)) = (2N_1 + N_0, 1+n), \, (2N_1 + N_0 -1, 2+n), \dots,
(N_1 + N_0 + j_2+1, N_1 - j_2 + n)
$$
(note that $N_1 - j_2 + n = N_1 + N_0 + j_2 + \nu_2 - 2$) and is thus equal to
$N_1-j_2$. For an exponent of $n=N_0 + 2(j_2 - 1) + \nu_2$ in the variable
$t_k$,
(3.14) gives that we require (3.15a) with $j_1$ replaced by $n$. When (3.16)
has its maximum number of solutions we see that (3.15a) doesn't have any
solutions. Furthermore, by decreasing the number  of solutions of (3.16) by
say $a_1$, we see that the number of solutions of (3.15a) can increase by no
more than $a_1$ (since for (3.15a) to have a solution we require
$P(2N_1 + k) \ge n + N_1 + 1$) and so we conclude that (3.12b) is true.

The validity of (3.12) implies that as a function of $\epsilon$, the l.h.s. of
(3.6) divided by the r.h.s. is bounded in the finite plane. Furthermore, using
the method given in the paragraph including (2.22) above, it is straightforward
to
show that for large-$|\epsilon|$ the r.h.s. of (3.6) has a reciprocal power
law decay with exponent (3.10), which is the same as the large-$|\epsilon|$
behaviour of the l.h.s. (recall the sentence including (3.10)). Hence, by
Liouville's theorem, both sides of (3.6) are the same functions of $\epsilon$,
up to a multiplicative function independent of $\epsilon$. This is the required
result.

\vspace{.2cm}
Let us now specify the dependence on $r$ of the function $A(N_0,N_1,r)$ in
(3.6). For this purpose we observe that when $\epsilon = r$, the l.h.s. of
(3.6) is independent of $r$. More explicitly,
\begin{eqnarray*}\lefteqn{
{\rm CT} \prod_{l=1}^{N_0} t_l^r\prod_{l=1}^{N_1}s_l^r
f(-t_1,\dots,-t_{N_0};-s_1,\dots,-s_{N_1};r)} \\
& = & {\rm CT}\prod_{l=1}^{N_0} (1+t_l)^{2r}\prod_{l=1}^{N_1} (1+s_l)^{2r}
f(-t_1,\dots,-t_{N_0};-s_1,\dots,-s_{N_1};0) \\
& = &{\rm CT} f(-t_1,\dots,-t_{N_0};-s_1,\dots,-s_{N_1};0)
\end{eqnarray*}
where the last line follows from the second last line after noting $f$ is a
homogeneous function of order 0.
For the r.h.s. of (3.6) to have this property, we require
$$
A(N_0,N_1,r) = B(N_0,N_1) \prod_{l=0}^{N_0-1} \Gamma (2r+1+l)
\prod_{l=0}^{N_1-1} \Gamma (2r+1+2l+N_0)
\eqno (3.18)
$$

The remaining unknown function $B(N_0,N_1)$ can be specified immediately from
the anlaytic result (3.1), or alternatively by using the general relationship
$$
D_1(N_1;N_0;\lambda,\lambda,\lambda) = D_1(N_1,N_0+1;0,0,\lambda)
\eqno (3.19)
$$
together with (2.25). We find
$$
B(N_0,N_1) =  \prod_{l=0}^{N_0-1}\Gamma (l+2) \prod_{j=0}^{N_1-1}(j+1)
\Gamma (2(j+1) + N_0)
\eqno (3.20)
$$
Substituting (3.20) in (3.18), then substituting the resulting expression in
the r.h.s. of (3.6), we obtain the exact evaluation of the integral in (3.6).
This exact evaluation agrees with the conjecture (3.3), and thus proves
the conjecture.

\vspace{.2cm}
\noindent
{\bf 3.2 General \mbox{\boldmath $\lambda$}}

\noindent
Using Morris's integral (1.3), the result (3.4) for $\lambda =1$, and the
analytic result (3.1) as guides, we conjecture that for general $\lambda$
\renewcommand{\theequation}{3.21}
\begin{eqnarray}
\lefteqn{D_1(N_1;N_0;a,b,\lambda)} \nonumber \\ & = & D(N_0;a,b,\lambda)
\prod_{j=0}^{N_1 - 1} {(j+1) \Gamma ((\lambda + 1)j+a+b+\lambda N_0+1)
\Gamma((\lambda + 1)(j+1)+\lambda N_0) \over
\Gamma(1 + \lambda)  \Gamma ((\lambda + 1)j+a+\lambda N_0+1)
\Gamma ((\lambda + 1)j+b+\lambda N_0+1)}\nonumber \\
\end{eqnarray}
where $D(N_0;a,b,\lambda)$ is given by (1.3). As well as being consistent with
theorems used in its formulation, (3.21) satisfies the general
relationship (3.19), and the consistency of the  large-$|\epsilon|$
 behaviour of both sides (recall
Section 2.3) can be checked as can  the independence on $a$ when $b=0$ (recall
the paragraph above
(3.18)).

\vspace{.2cm}
\noindent
{\bf 3.3 Normalization of the harmonic well wavefunction for \mbox{\boldmath
$p=1$}}

\noindent
{}From the conjecture (3.21), it is possible to deduce the value of the
integral
\renewcommand{\theequation}{3.22}
\begin{eqnarray}
\lefteqn{G_p(N_1, \dots,N_p;N_0;\lambda)} \nonumber \\
& & := \left ( \prod_{\alpha = 1}^p \prod_{j=1}^{N_\alpha}
\int_{-\infty}^\infty
dx_j^{(\alpha)} \right ) \left (\prod_{j'=1}^{N_0} \int_{-\infty}^\infty
dy_{j'} \right ) |\psi_0^{(h)}(\{x_j^{(\alpha)}\}_{\alpha = 1,\dots,p \atop
j=1,\dots,N_\alpha},
\{y_{j'}\}_{j'=1,\dots,N_0})|^2 \nonumber \\
\end{eqnarray}
where $\psi_0^{(h)}$ is given by (1.4), in the case $p=1$.

By setting $a=b$ in (2.24) and changing variables $x_j^{(1)} \mapsto
x_j^{(1)}/2 \pi a$, $y_l \mapsto y_l/2 \pi a$, we see that for $a \rightarrow
\infty$
\renewcommand{\theequation}{3.23}
\begin{eqnarray}
\lefteqn{D_1(N_1;N_0;a,a,\lambda)} \nonumber \\
& & \sim \left ({1 \over 2 \pi } \right )^{N_1 + N_0} \left ( { 1 \over a}
\right )^{(\lambda + 1)N_1(N_1-1) + \lambda N_0N_1+ \lambda N_0(N_0-1)}
2^{2 a (N_1 + N_0}
G_1(N_1;N_0;\lambda)
\end{eqnarray}
On the other hand, with $a=b$ it is straightforward to obtain the large-$a$
behaviour of the conjectured evaluation (3.21) of $D_1$ by using Stirling's
formula. Comparison with (3.23) then gives
$$
G_1(N_1;N_0;\lambda) = ( 2 \pi)^{N_1 + N_0)/2} \prod_{j=0}^{N_0-1}
{\Gamma ( 1 + \lambda (j+1)) \over \Gamma ( 1 + \lambda)}
\prod_{k=0}^{N_1-1} {\Gamma ((\lambda + 1)(k+1) + \lambda N_0) \over \Gamma(1+
\lambda)}
\eqno (3.24)
$$
as the conjectured evaluation of $G_1$ (in the case $\lambda = 1$, since
we have proved (3.21), we also have proved (3.24)).

\vspace{.5cm}

\noindent
{\bf 4. THE GENERAL \mbox{\boldmath $p$} cases}

\vspace{.2cm}

\noindent
{\bf 4.1 The case \mbox{\boldmath $p=2$} }

\noindent
With $p=2$ we have the general relations
$$
D_2(0,N_2;0;a,b,\lambda) = D(N_2;a,b,\lambda + 1)
\eqno (4.1)
$$
where the r.h.s. is given by Morris's integral (2.3), and
$$
D_2(1,N_2;0;0,0,\lambda) =D_1(N_2;1;0,0\lambda)
\eqno (4.2)
$$
where the r.h.s. is given by the conjecture (3.21).

These formulas provide analytic data for the cases $N_0=0$ and 1. For $N_1=2$
and 3, and with
$\lambda =1, \, a=b=0$, numerical data was obtained. By sequentially increasing
$N_2$, the data was seen to fit the forms
\renewcommand{\theequation}{4.3}
\begin{eqnarray}
D_2(2,N_2;0;0,0,1) &= &{16 \over 3} \prod_{j=1}^{N_2} j (2j+1), \qquad N_2 \ge
1 \\
D_2(3,N_2;0;0,0,1) &= & 70 \prod_{j=1}^{N_2} j (2j+2), \qquad N_2 \ge 2
\end{eqnarray}
The results (4.1)-(4.4) suggest that
\renewcommand{\theequation}{4.5}
\begin{eqnarray}
D_2(N_1,N_2;0;0,0,1) & = & g(N_1) \prod_{j=1}^{N_2} j (2j - 1 + N_1) \nonumber
\\
& = & g(N_1) \prod_{j=0}^{N_2 - 1} {(j+1) \Gamma (2(j + 1) + N_1) \over
\Gamma ( 2j + 1 + N_1) } , \qquad N_2 \ge N_1 - 1
\end{eqnarray}

For general $a,b$ and $\lambda$ inspection of (4.5) and use of (4.1) and (1.3)
suggest the same ansatz used in (3.3) for $D_1(N_1;N_0;a,b,1)$:
$$
D_2(N_1,N_2;0;a,b,\lambda) = f(N_1,a,b) A(N_1,N_2;a,b,\lambda)
\eqno (4.6a)
$$
where
$$
A(N_1,N_2;a,b,\lambda) =
\prod_{j=0}^{N_2 - 1} {(j+1) \Gamma ((\lambda + 1)j+a+b+N_1+1)
\Gamma((\lambda + 1)(j+1)+N_1) \over
  \Gamma ((\lambda + 1)j+a+N_1+1)\Gamma ((\lambda + 1)j+b+N_1+1)}
\eqno (4.6b)
$$
valid for
$$
N_2 \ge N_1 - 1
\eqno (4.6c)
$$
The restriction (4.6c) is the key distinguishing feature between (3.3) and
(4.6a).
The function $f(N_1,a,b)$ can be determined from the symmetry relation
$$
D_2(N,N-1;0;a,b,\lambda) = D_2(N-1,N;0;a,b,\lambda)
$$
which gives the difference equation
$$
f(k;a,b,\lambda) A(k,k-1;a,b,\lambda) = f(k-1;a,b,\lambda) A(k-1,k;a,b,\lambda)
$$
This difference equation has solution
$$
f(N_1;a,b,\lambda) = \prod_{k=1}^{N_1} {A(k-1,k;a,b,\lambda) \over
A(k,k-1;a,b,\lambda)}
\eqno (4.7)
$$
where we have used the fact that
$$
f(0;a,b,\lambda) = 1
$$
which follows by choosing $N_1 = N_2 = 0$ in (4.6a). Substituting (4.7) in
(4.6a)
gives the conjectured evaluation of $D_2(N_1,N_2;0;a,b,\lambda)$.

\vspace{.2cm}
\noindent
{\bf 4.2 The general case}

\noindent
Guided by (4.6a) and (3.3), for general $p$ and $N_0$ we conjecture that
$$
D_p(N_1, \dots,N_p;N_0;a,b,\lambda)  =  f_{p-1}(N_1,\dots,N_{p-1};N_0;
a,b,\lambda) A_p(N_1, \dots, N_p; N_0;a,b,\lambda)
\eqno (4.8a)
$$
where
\renewcommand{\theequation}{4.8b}
\begin{eqnarray}\lefteqn{A_p(N_1, \dots, N_p; N_0;a,b,\lambda)} \nonumber \\
&  & =
\prod_{j=0}^{N_p - 1} {(j+1) \Gamma ((\lambda + 1)j+a+b+\lambda
\sum_{j=0}^{p-1}
 N_j+1)
\Gamma((\lambda + 1)(j+1)+\lambda \sum_{j=0}^{p-1}N_j) \over
\Gamma(1 + \lambda)  \Gamma ((\lambda + 1)j+a+\lambda \sum_{j=0}^{p-1}N_j +1)
\Gamma ((\lambda + 1)j+b+\lambda\sum_{j=0}^{p-1}N_j +1)} \nonumber \\
\end{eqnarray}
and $N_p \ge N_j - 1 \, (j=1, \dots, p-1)$. To calculate $f_{p-1}$ we make the
ordering
$$
N_j \ge N_{j-1} \qquad (j=2, \dots, p)
\eqno (4.8c)
$$
and use the symmetry relation
$$
D_p(N_1, \dots,N_{k-2},N-1,N,N_{k+1}, \dots,N_p;N_0;a,b,\lambda) \hspace{3cm}
$$
$$
= D_p(N_1, \dots,N_{k-2},N,N-1,N_{k+1}, \dots,N_p;N_0;a,b,\lambda)
\eqno (4.9)
$$
for $k=2, \dots, p$. From (4.9) and the initial condition
$$
D_p(0, \dots,0;N_0;a,b,\lambda) = D(N_0;a,b,\lambda)
$$
where $ D(N_0;a,b,\lambda)$ is given by (2.3), we obtain the recurrence
equations
\renewcommand{\theequation}{4.10a}
\begin{eqnarray}\lefteqn{f_{k-1}(N_1, \dots, N_{k-1};N_0;a,b,\lambda)}
\nonumber \\
& & = A_{k-1}(N_1, \dots, N_{k-1}; N_0;a,b,\lambda)
f_{k-2}(N_1, \dots, N_{k-2};N_0;a,b,\lambda)
\end{eqnarray}
where
$$
A_{k-1}(N_1, \dots, N_{k-1}; N_0;a,b,\lambda)
:= \prod_{j=1}^{N_{k-1}}{A_k(N_1, \dots,N_{k-2},j-1,j;N_0;a,b,\lambda)
\over A_k(N_1, \dots,N_{k-2},j,j-1;N_0;a,b,\lambda)}
\eqno (4.10b)
$$
and
$$
f_0(N_0;a,b,\lambda) = D(N_0;a,b,\lambda)
\eqno (4.10c)
$$
Taken in the order $k=p,p-1,\dots,2$ these equations explicitly determine
$f_{p-1}$ and thus $D_p$.

 For example, with $p=3$ we obtain
\renewcommand{\theequation}{4.11a}
\begin{eqnarray}\lefteqn{D_3(N_1,N_2,N_3;N_0;a,b,\lambda)} \nonumber \\
&  & = D(N_0;a,b,\lambda) A_3(N_1,N_2,N_3;N_0;a,b,\lambda) \prod_{j=1}^{N_2}
{A_3(N_1,j-1,j;N_0;a,b,\lambda) \over A_3(N_1,j,j-1;N_0;a,b,\lambda)} \nonumber
\\
& & \times \prod_{k=1}^{N_1}{A_3(k-1,k-1,k;N_0;a,b,\lambda)
 \over A_3(k-1,k,k-1;N_0;a,b,\lambda)}
\prod_{j=1}^{k-1}{A_3(k-1,j-1,j;N_0;a,b,\lambda)A_3(k,j,j-1;N_0;a,b;\lambda)
\over
 A_3(k-1,j,j-1;N_0;a,b,\lambda)
A_3(k,j-1,j;a,b,\lambda) } \nonumber \\
\end{eqnarray}
where
$$
N_p \ge N_{p-1} - 1 , \qquad p=2,3
\eqno (4.11b)
$$
Note that this agrees with (4.6a) when $N_1=N_0=0$. Also, we have made the
exact numerical evaluations
$$
D_3(1,2,2;0;0,0,1) = 720 \qquad {\rm and} \qquad D_3(2,2,2;0;0,0,1) = 10,080
$$
and found agreement with (4.11).

\vspace{.2cm}
\noindent
{\bf 4.3 Normalization of the harmonic well wavefunction in the general case}

\noindent
Analogous to (4.6a), for the integral (3.22) we conjecture
$$
G_p(N_1, \dots,N_p;N_0;\lambda) = g_{p-1}(N_1,\dots,N_{p-1};N_0;\lambda)
B_p(N_1,\dots,N_p;N_0;\lambda)
\eqno (4.12a)
$$
where
$$
B_p(N_1, \dots,N_p;N_0;\lambda) := (2 \pi)^{N_p/2} \prod_{j=0}^{N_p - 1}
{(j+1) \over \Gamma (1 + \lambda)} \Gamma ((\lambda + 1)(j+1) + \lambda
\sum_{j=0}^{p-1} N_j )
\eqno (4.12b)
$$
With the ordering (4.8c), and assuming the analogue of the symmetry relation
(4.9), the conjecture (4.11a) gives the recurrence equations
$$
g_{k-1}(N_1, \dots,N_{k-1};N_0;\lambda) = B_{k-1}(N_1,
\dots,N_{k-1};N_0;\lambda)
g_{k-2}(N_1, \dots,N_{k-2};N_0;\lambda)
\eqno (4.13a)
$$
where
$$
B_{k-1}(N_1, \dots,N_{k-1};N_0;\lambda) : = \prod_{j=1}^{N_{k-1}}
{B_k (N_1, \dots,N_{k-2},j-1,j;N_0;\lambda) \over
B_k (N_1, \dots, N_{k-2},j,j-1;N_0;\lambda)}
\eqno (4.13b)
$$
and
$$
g_0(N_0;\lambda) = (2 \pi)^{N_0 /2} \prod_{j=0}^{N_0 - 1}
{ (j+1) \Gamma (\lambda (1 + j)) \over \Gamma (1 + \lambda)}
\eqno (4.13c)
$$
which when taken in the order $k=p,p-1, \dots ,2$ explicitly determine
$g_{p-1}$ and thus $G_p$.

\vspace{.5cm}
\noindent
{\bf 5. SUMMARY}

\noindent
The objective of this paper has been to provide the exact evaluation of the
trigonometric integral (2.24). This integral is a generalization of Morris's
integral (1.3) (which is equivalent to Selberg's integral (1.1)), and
includes as a special case the normalization of the multicomponent wavefunction
(1.2). We have been partially successful in this task in that (4.8) and (4.10)
provide the conjectured exact evaluation of (2.24) expressed in the forms of
recurrence equations. Moreover (4.8) provides a conjecture for a specific
functional property of the integrals (2.24), from which their exact
evaluation follows.

In Section 2.3 we have also provided a new proof of Morris's integral, which
was used in Section 3.1 to prove the conjectured evaluation of
$D_1(N_1;N_0;a,b,1)$. However we have not been successful in providing a
proof in the general case, which we leave as an open problem.

\pagebreak
\noindent
{\bf References}
\begin{description}
\item[][1] P. Pechukas, Phys. Rev. Lett. {\bf 51} (1983) 943;
\item[]\hspace*{2ex} T. Yukawa, Phys. Lett. A {\bf 116} (1986) 227.
\item[][2] A. Polychronakos, Nucl. Phys.  B {\bf 234} (1989) 597;
\item[]\hspace*{2ex} F.D.M. Haldane, in {\it Proceedings of the 16th Taniguchi
Symposium} eds. A. Okiji and N. Kawakami (Springer-Verlag, 1994).
\item[][3] Z.N.C. Ha, Phys. Rev. Lett. {\bf 73} (1994) 1574.
\item[][4] P.J. Forrester, Nucl. Phys. B {\bf 388} (1992) 671, Nucl. Phys. B
{\bf 416} (1994) 377;
\item[]\hspace*{2ex} F. Lesage, V. Pasquier and D. Serban, hep-th/9405008
submitted Nucl. Phys. B.
\item[][5] A. Selberg, Norsk. Mat. Tidsskr. {\bf 26} (1944) 71;
\item[]\hspace*{2ex} K. Aomoto, SIAM J. Math. Anal. {\bf 18} (1987) 545;
\item[]\hspace*{2ex} G.W. Anderson, Forum Math. {\bf 3} (1991) 415.
\item[][6] Z.N.C. Ha and F.D.M. Haldane, Phys. Rev. B {\bf 46} (1992) 9359;
\item[]\hspace*{2ex} N. Kawakami, Phys. Rev. B {\bf 46} (1992) 3191;
\item[]\hspace*{2ex} Y. Kato and Y. Kuramoto, cond-mat/9409038      preprint.
\item[][7] Y. Kuramoto and H. Yokoyama, Phys. Rev. Lett. {\bf 67} (1991) 1338.
\item[][8] W.G. Morriss, Ph.D. thesis, University of Wisconsin at Madison
(1982).
\item[][9] P. Di Francesco, M. Gaudin, C. Itzykson and F. Lesage, Int. J. of
Mod.
Phys. A {\bf 9} (1994) 4247.
\item[][10] D.M. Bressoud and I.P. Goulden, Commun. Math. Phys. {\bf 110}
(1987)
287.
\item[][11] P.J. Forrester and B. Jancovici, J. Physique Lett. {\bf 45} (1984)
L583.

\end{description}

\end{document}